\documentclass[prb,twocolumn,showpacs,floats,eqsecnum,amsmath,amssymb,footinbib]{revtex4}
\usepackage[dvips]{graphicx}

\begin{document}



\title{Formation of clusters in the ground state of the $t-J$ model
on a two leg ladder}

\author{A. Fledderjohann$^{1}$, A. Langari$^{2}$ and K.-H. M\"utter$^{1}$}

\affiliation{$^1$Physics Department, University of Wuppertal, 42097 Wuppertal,
Germany}
\affiliation{
$^2$Institute for Advanced Studies in Basic Sciences,
Zanjan 45195-159, Iran}
\date{\today}


\begin{abstract}
\leftskip 2cm
\rightskip 2cm

We investigate the ground state properties of the $t-J$ model on a two leg
ladder with anisotropic couplings ($t,\alpha=J/t$) along rungs
and ($t',\alpha'=J'/t'$) along legs.
We have implemented a cluster approach based on 4-site plaqettes.
In the strong asymmetric cases
$\alpha/\alpha'\ll 1$ and $\alpha'/\alpha\ll 1$ the ground state energy is
well described by plaquette clusters with charges $Q=2,4$. The interaction
between the clusters favours the condensation of plaquettes with maximal
charge -- a signal for phase separation. The dominance of $Q=2$ plaquettes
explains the emergence of tightly bound hole pairs.
We have presented the numerical  results of exact diagonalization to
support our cluster approach.
\end{abstract}

\pacs{71.10.Fd,71.27.+a,75.10.Jm}

\maketitle



\section{Introduction: Motivation and questions of interest}
\setcounter{equation}{0}

After the discovery of high $T_c$ superconductors\cite{bednorz86} -- almost twenty
years ago -- models of strongly correlated electron systems doped
with holes have attracted much interest. The $t-J$ model in two
dimensions \cite{anderson87,dagotto92,emery90,putikka92,fehske91,
valenti92,hellberg95_00,calandra98}
and on ladders\cite{rommer00,martins99}
has been studied intensively in order
to understand the behaviour of mobile holes in an antiferromagnetic
background. Although some exact results in the special form of interactions
exist\cite{bose94,frahm99}
the ground state and low energy excitations on the $t-J$ ladder
are not known exactly.

The generic mechanisms which explain the most striking features are
of special interest, namely:
\begin{itemize}
\item[i)]
the opening of a charge transfer gap\cite{zaanen85},
\item[ii)]
the spatial separation of phases with hole rich and hole poor
domaines\cite{jorgensen88}.
\end{itemize}
Signatures of these features can be seen already in the ground
state energy per site
\begin{eqnarray}
\varepsilon(\rho)=\frac{E_{G}}{N}, &\hspace{1.0cm} & \rho=\frac{Q_{\mbox{tot}}}{N},
\end{eqnarray}
where $E_G$, $N$, $Q_{\mbox{tot}}$ are the ground state energy,
the number of sites and the total charge, respectively.
\begin{itemize}
\item[i)]
A discontinuity in the first derivative, i.e. chemical potential,
\begin{eqnarray}
\mu(\rho) & = & \frac{d\varepsilon}{d\rho},\\
\mu(\rho) & = & \left\{\begin{array}{ll}
\mu_- & \rho\rightarrow\rho_0-0\\
\mu_+ & \rho\rightarrow\rho_0+0,
\end{array}\right.
\end{eqnarray}
signals the opening of a gap.

The inverse function
\begin{eqnarray}
\rho(\mu) & = & \rho_0\quad\mbox{for\quad}\mu_-\leq\mu\leq\mu_+,
\end{eqnarray}
develops a plateau with a width
\begin{eqnarray}
\Delta & = &  \mu_+ -\mu_-,
\end{eqnarray}
which is related to the charge transfer gap.
This is quite analogous to the plateaux in the magnetization
curve $M=M(B)$, which
are related to the spin gap.



\item[ii)]
A linear behaviour of $\varepsilon(\rho)$ in some interval,
$\rho_1\leq\rho\leq\rho_2$:
\begin{equation}
\varepsilon_L(\rho) = \frac{1}{\rho_2-\rho_1}\left[
\varepsilon(\rho_1)(\rho_2-\rho)+\varepsilon(\rho_2)
(\rho-\rho_1)\right], \label{eps_lin}
\end{equation}
signals the spatial separation of two phases: the first one with
charge density $\rho_1$, the second one with $\rho_2$.

For demonstration, let us consider a two leg ladder, which we
divide into two parts with sites $N_j$, charges $Q_j$,
Hamiltonians $H_j$, ground state wave functions $\psi(\rho_j,N_j)$
and ground state energies $\varepsilon(\rho_j)N_j(\rho)$ as
illustrated in Fig. \ref{fig1_in}.

\begin{figure}[ht!]
\centerline{\hspace{0.0cm}\includegraphics[width=7.5cm,angle=0]{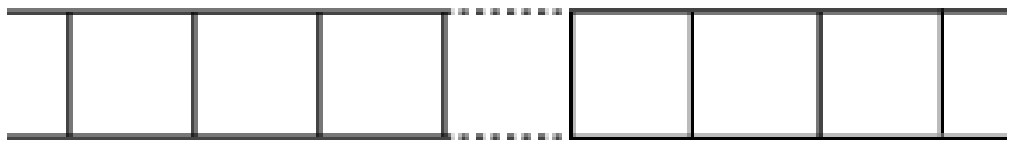}
\hspace{-1.0cm}}

\vspace{-0.4cm}
\begin{eqnarray}
H_1(N_1(\rho))\hspace{0.4cm} & & \hspace{0.5cm}H_{12}\hspace{1.0cm}
H_2(N_2(\rho))\nonumber\\[22pt]
\hspace{0.5cm}
N_1(\rho)=\frac{\rho_2-\rho}{\rho_2-\rho_1}N  ,\hspace{-0.1cm} & &\hspace{1.5cm}
N_2(\rho)=\frac{\rho-\rho_1}{\rho_2-\rho_1}N\hspace{0.0cm}\nonumber\\[9pt]
Q_1=N_1(\rho)\rho_1  ,\hspace{-0.1cm} & &\hspace{1.5cm}
Q_2=N_2(\rho)\rho_2\nonumber\\[9pt]
\psi(\rho_1,N_1(\rho)),\hspace{-0.1cm} & &\hspace{1.5cm}
\psi(\rho_2,N_2(\rho))\nonumber\\[9pt]
\varepsilon(\rho_1)N_1(\rho) ,\hspace{-0.1cm} & &\hspace{1.5cm}
\varepsilon(\rho_2)N_2(\rho)\nonumber
\end{eqnarray}
\caption{Subdivision of a two leg ladder
}
\label{fig1_in}
\end{figure}

\vspace{0.0cm}



The Hamiltonian of the whole ladder
\begin{eqnarray}
H & = & H_1(N_1(\rho))+
H_2(N_2(\rho))+
H_{12}\label{H_tot},
\end{eqnarray}
contains in addition an interaction term, which is mediated via the two
(dashed) links connecting the subladders.
Each bond in Fig.\ref{fig1_in} corresponds to the usual $t-J$
Hamiltonian, composing of an electron hopping term $t(c_{i,\sigma}^{\dagger}c_{j,\sigma}
+c_{j,\sigma}^{\dagger}c_{i,\sigma})$
and an exchange interaction $J(S_i\cdot S_j-n_in_j/4)$.
The elimination of doubly occupied states has been also imposed.
The indices $i,j$
refer to nearest neighbour sites, $\sigma$ to electron spin
and the couplings  on legs are defined as ($t', J'$).

The product ansatz:
\begin{eqnarray}
\psi(\rho,N) & = & \psi(\rho_1,N_1(\rho))\psi(\rho_2,N_2(\rho)),
\label{var_ansatz}
\end{eqnarray}
describes a state with charge density $\rho$ and two spatially
separated phases, the first one with charge density $\rho_1$
and $N_1(\rho)$ sites and the second with charge density $\rho_2$
and $N_2(\rho)$ sites.

If we estimate the expectation value of the Hamiltonian (\ref{H_tot})
with the variational ansatz (\ref{var_ansatz}), we get an upper bound
\begin{eqnarray}\label{upper_bound}
\varepsilon(\rho) & \leq & \varepsilon_L(\rho)
\quad\mbox{for\quad}\rho_1\leq\rho\leq\rho_2,
\end{eqnarray}
in terms of the right-hand side of (\ref{eps_lin}).

Note that the interaction term
\begin{eqnarray}
 & & \frac{1}{N}\langle\psi(\rho) |H_{12}|\psi(\rho)\rangle,\nonumber
\end{eqnarray}


does not survive in the thermodynamical limit
$N\rightarrow\infty$.
Since (\ref{upper_bound}) is a strict upper bound in the
thermodynamical limit for any interval
\begin{eqnarray}
\rho_1 & \leq & \rho\leq\rho_2,\nonumber
\end{eqnarray}
we can conclude that $\varepsilon(\rho)$ is a convex function of $\rho$.
The product ansatz (\ref{var_ansatz}) with the two separated phases represents
the true ground state if the upper bound (\ref{upper_bound})
sharply holds. In this case, we derive from (\ref{eps_lin})
a constant chemical potential:

\begin{eqnarray}
\mu=\frac{d\varepsilon}{d\rho} & = & \frac{\varepsilon(\rho_2)-
\varepsilon(\rho_1)}{\rho_2-\rho_1}=\mu_0\quad
\mbox{for}\quad\rho_1\leq\rho\leq\rho_2,\hspace{-0.8cm}\nonumber\\
\end{eqnarray}
which corresponds to a discontinuity in the inverse function
\begin{eqnarray}
\rho(\mu) & = & \left\{\begin{array}{ll}
\rho_1 & \mbox{for}\quad\mu\rightarrow\mu_0+0\\
\rho_2 & \mbox{for}\quad\mu\rightarrow\mu_0-0.
\end{array}\right.
\end{eqnarray}
\end{itemize}

It is the purpose of this paper, to demonstrate that the generic
mechanisms, which lead to gaps and phase separations, are
intimately related with the formation of clusters.
On ladder systems the formation of clusters is prescribed in a
natural way by the ladder geometry.

It is plausible to start with the simplest clusters, defined by the rungs.
In the limit of vanishing hopping parameter $t'$ along the legs,
the system decouples into a product
of rung eigenstates.
This limit has been studied intensively in the literature
under various notations like ''local rung approximation''
\cite{riera99} or ''bond operator theory'' \cite{park01}.
In Ref. \cite{fl03}, we studied first order corrections
in the leg hopping parameter $t'$ and compared  perturbative
results with exact diagonalizations on a $2\times 8$ ladder
for parameter values
\begin{eqnarray}
t=1, & \alpha=J=0.5,\quad & \\
t'=0.1,0.2,0.3, & \alpha'=J'/t'=2.7 \;. & \nonumber
\label{coupl_parm_1}
\end{eqnarray}
Concerning the quality of the perturbation
expansion based on the local rung approximation we found
\begin{itemize}
\item
good agreement in the regime $0\leq\rho\leq 1/2,$
\item
failure in the regime $1/2\leq\rho\leq 1.$
\end{itemize}
We do not think that higher orders in perturbation expansion with
local rungs will improve the situation in the second regime.
Instead we are convinced the starting point -- i.e. the clusters
which define the zeroth order perturbation theory -- has to be
changed.

In this paper we intend to demonstrate on the two leg ladder
with anisotropic couplings, how the appropriate clusters which
define the zeroth order perturbation theory are to be found.
We start in Section \ref{sec2} with an analysis of the exact
ground state energy per site $\varepsilon(\rho)$ on a $2\times 8$
ladder [cf. Fig. \ref{nfig1}(a),(b)] which turns out to be
almost linear in the charge density $\rho$ for an appropriate
choice of the system parameters (Eqs.(\ref{tuning})(a) and (\ref{tuning})(b)).
As explained above the linearity in $\rho$ indicates phase
separation into two clusters with charge densities
\begin{eqnarray}
\rho_1=0\quad\rho_2=\frac{1}{2} & \mbox{\,\,for\,\,} & 0\leq\rho\leq \frac{1}{2},\\
\rho_1=\frac{1}{2}\quad\rho_2=1 & \mbox{\,\,for\,\,} & \frac{1}{2}\leq\rho\leq 1.
\end{eqnarray}
The dependence of the ground state energy per site (\ref{eps_lin})
on the system parameters
is very well reproduced by the plaquette ground state energies
$E^{(p)}(Q)$ with plaquette charges $Q=2,4$.

%

In Section \ref{sec3} we present the perturbation expansion
based on plaquette clusters.
First order corrections, which describe the interaction
between neighbouring plaquettes, favour the clustering of
plaquettes with charges $Q=4$ -- a first indication of
phase separation.

In Section \ref{sec4} we investigate under which circumstances
plaquette clusters with odd charges $Q=1$ and $Q=3$ are
suppressed energetically.

In the low doping case $\rho>3/4$ the dominance of $Q=2$ and
suppression of $Q=3$ plaquettes in the ground state explains
the emergence of tightly bound hole pairs (Ref. \cite{siller01})
for an appropriate choice of the rung and leg couplings.
Finally, a discussion on our results will be presented.

\section{Approximate linear charge density dependence of the ground
state energy\label{sec2}}

\begin{figure}[ht!]
\centerline{\includegraphics[height=8.0cm,angle=-90]{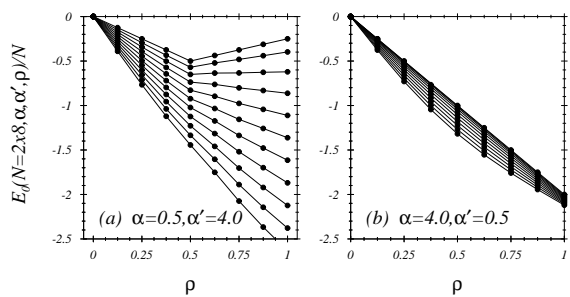}}
\caption{Lanczos results for energies per site of a $N=2\times 8$ 
$t-J$ ladder for the parameters (\ref{tuning})(a),(b). For both 
shown cases $t'$ 
increases from top to bottom as $t'=0,0.1,\ldots 1.0$.}
\label{nfig1}
\end{figure}

We start in Figs. \ref{nfig1}(a), \ref{nfig1}(b) with Lanczos
results of the ground state energy per site in the $t-J$ model
with anisotropic couplings $t,\,\alpha=J/t$ (throughout this paper
we have chosen $t\equiv 1$), $t',\,\alpha'=J'/t'$
for the rungs and legs respectively on a $2\times 8$ ladder:
\begin{eqnarray}
\mbox{(a)}\quad\alpha=0.5,\,\,\alpha'=4.0, & & t'=0,0.1,\ldots, 1.0,\nonumber\\
\mbox{(b)}\quad\alpha=4.0,\,\,\alpha'=0.5, & & t'=0,0.1,\ldots, 1.0\,.\nonumber\\
 & &\label{tuning}
\end{eqnarray}

There is a pronounced difference in the $\rho$-dependence of the
ground state energies in Fig. \ref{nfig1}(a) and Fig. \ref{nfig1}(b),
respectively.
In Fig. \ref{nfig1}(a) we observe a discontinuity in the slope 
$\mu(\rho)=d\varepsilon/d\rho$ at $\rho=1/2$, which generates a
plateau in the charge density $\rho(\mu)$ as function of the
chemical potential
\begin{eqnarray}
\rho(\mu)=\frac{1}{2} & \quad\quad & \mu_-\leq\mu\leq\mu_+\,.
\end{eqnarray}
The plateau width
\begin{eqnarray}
\Delta & = & \mu_+-\mu_-
\end{eqnarray}
shrinks with increasing values of $t'$. Note also that the chemical
potential $\mu(\rho)=d\varepsilon/d\rho$ vanishes for $\rho>1/2$
for a specific value of $t'=t_0'(\alpha,\alpha')$. This feature
will be discussed in Section \ref{sec4}.

Let us next turn to the ground state energy $\varepsilon(\rho)$
in the domain (\ref{tuning})(b) shown in Fig. \ref{nfig1}(b).
At $t'=0$, there is no discontinuity in the slope -- i.e. no
plateau in $\rho(\mu)$ -- at $\rho=1/2$.
The variation with the leg hopping parameter $t'$ is much smaller
than in case $(a)$. For $t'>0$, $\varepsilon(\rho,t')$ is only
approximately linear in the two subintervals $0\leq\rho\leq 1/2$ 
and $1/2\leq\rho\leq 1$ with different slopes $\mu_-$ and $\mu_+$.
The difference $\Delta=\mu_+-\mu_-$ increases with $t'$. Deviations
from linearity are convex as predicted by (\ref{upper_bound}). 

The product ansatz (\ref{var_ansatz}) describes a system with two phases:

For $0\leq\rho\leq 1/2$ with a ground state energy
\begin{eqnarray}
\varepsilon(\rho,t') & = & \varepsilon(0,t')(1-2\rho)+
\varepsilon(1/2,t')2\rho
\label{eps0_lt}
\end{eqnarray}
there is a phase with charge density $\rho_1=0$ in the first
part of the ladder and a second phase with charge density $\rho_2=1/2$
in the second part. Phase separation occurs at $N_1(\rho)=(1-2\rho)N$.

For $1/2\leq\rho\leq 1$ with a ground state energy
\begin{eqnarray}
\varepsilon(\rho,t') & = & 2\varepsilon(1/2,t')(1-\rho)+
\varepsilon(1,t')(2\rho-1)
\label{eps0_gt}
\end{eqnarray}
the two phases in the first and second part of the ladder
have charge densities $\rho_1=1/2$ and $\rho_2=1$, respectively.
Here the phase separation occurs at $N_1(\rho)=2(1-\rho)N$.

\begin{figure}[ht!]
\centerline{\includegraphics[height=8.0cm,angle=-90]{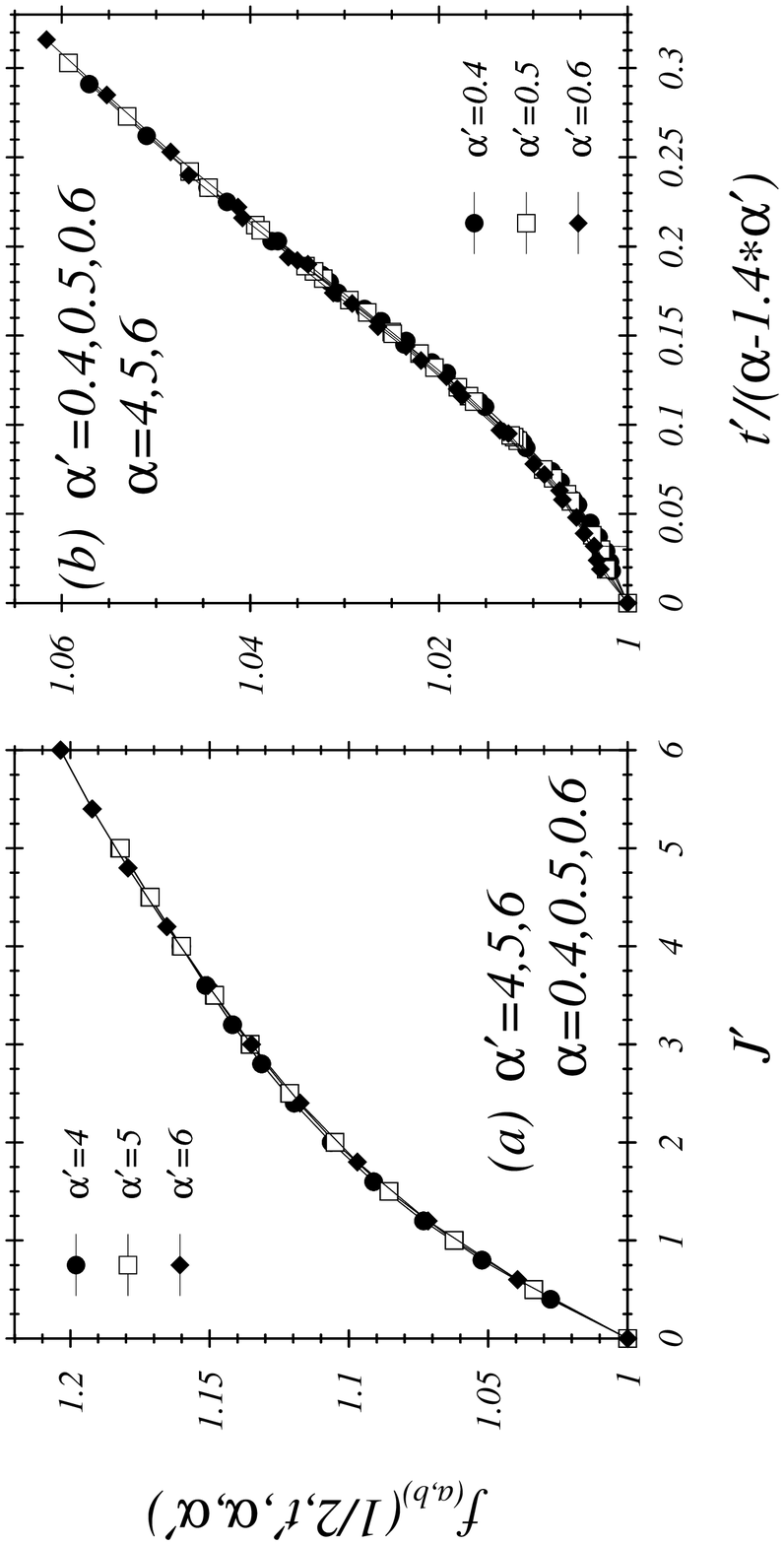}}
\caption{Correction factor $f(1/2,t',\alpha ,\alpha')$ of
(\ref{frac_eps}) for a $2\times 8$ $t-J$ ladder with
parameters (\ref{coupl_parm_2})(a),(b) and $\rho=1/2$}
\label{nfig2}
\end{figure}

For $0\leq\rho\leq 1/2$ the dependence of $\varepsilon(\rho,t')$
(\ref{eps0_lt}) on the parameters $t',\alpha,\alpha'$ only enters
via $\varepsilon(1/2,t',\alpha,\alpha')$. Results on a
$2\times 8$ ladder are shown for this quantity in Figs. \ref{nfig2}(a) and
\ref{nfig2}(b) with the parameter choices
\begin{eqnarray}
\mbox{(a)}\,\,\alpha=0.4,0.5,0.6,\,\alpha'=4.0,5.0,6.0, & &
\hspace{-0.2cm}t'=0,0.1,\ldots, 1.0\,,\nonumber\\
\mbox{(b)}\,\,\alpha=4.0,5.0,6.0,\,\alpha'=0.4,0.5,0.6, & &
\hspace{-0.2cm}t'=0,0.1,\ldots, 1.0\,.\nonumber\\
 & &\label{coupl_parm_2}
\end{eqnarray}

It turns out that the whole dependence on the parameters
$t',\alpha,\alpha'$:
\begin{eqnarray}
\varepsilon(1/2,t',\alpha ,\alpha')=\hspace{4.0cm} & &\nonumber\\
\frac{1}{4}E^{(p)}(2,t',\alpha ,\alpha')\cdot f(1/2,t',\alpha ,\alpha')
 & & \label{frac_eps}
\end{eqnarray}
is correctly reproduced by the ($2\times 2$)-plaquette ground state energies
$E^{(p)}(Q,t',\alpha ,\alpha')$ with charge $Q=2$ up to a
correction factor $f(1/2,t',\alpha ,\alpha')$. The physical
interpretation of $f$ is given below.

As is shown in Appendix \ref{appendix_a} [cf. (\ref{E0_p})-(\ref{E_p_b})],
we have different ground states in the regimes (\ref{coupl_parm_2})(a) and
(\ref{coupl_parm_2})(b) for plaquettes with charge $Q=2$.
The ground state energies follow from the lowest eigenvalues
of the $3\times 3$ matrix (\ref{Q2_3x3}), which are computed
in a perturbation expansion with zeroth order contributions
(\ref{E_p_a}) and (\ref{E_p_b}) for the regimes
(\ref{coupl_parm_2})(a) and (\ref{coupl_parm_2})(b).

\begin{eqnarray}
\mbox{(a)}\quad E_{(a)}^{(P)}(2,t',\alpha ,\alpha') & = & -\frac{1}{2}\left(J'+\sqrt{
J'^2+16}\right)\,,\quad\nonumber\\
\mbox{(b)}\quad E_{(b)}^{(P)}(2,t',\alpha ,\alpha') & = & -\frac{1}{2}\left(\alpha+
\sqrt{\alpha^2+16t'^2}\right)\,.\quad\nonumber\\
 & & \label{Ep}
\end{eqnarray}
In Figs. \ref{nfig2}(a) and \ref{nfig2}(b) we have plotted the
correction factors
in (\ref{frac_eps}):
\begin{eqnarray}
\mbox{(a)}\quad f_{(a)}(1/2,t',\alpha ,\alpha') & = & f_{(a)}(1/2,
J'=\alpha't')\,,\quad\nonumber\\
\mbox{(b)}\quad f_{(b)}(1/2,t',\alpha ,\alpha') & = & f_{(b)}(1/2,
t'/(\alpha-1.4\alpha'))\,.\quad\nonumber\\
 & & \label{f_ab}
\end{eqnarray}
versus the ``scaling variables'' $J'=\alpha't'$ and $t'/(\alpha-1.4\alpha')$,
respectively.
Note, that all data points for  (\ref{coupl_parm_2})(a) and
(\ref{coupl_parm_2})(b) almost coincide if we use the scaling variables. 

\begin{figure}[ht!]
\centerline{\includegraphics[height=8.0cm,angle=-90]{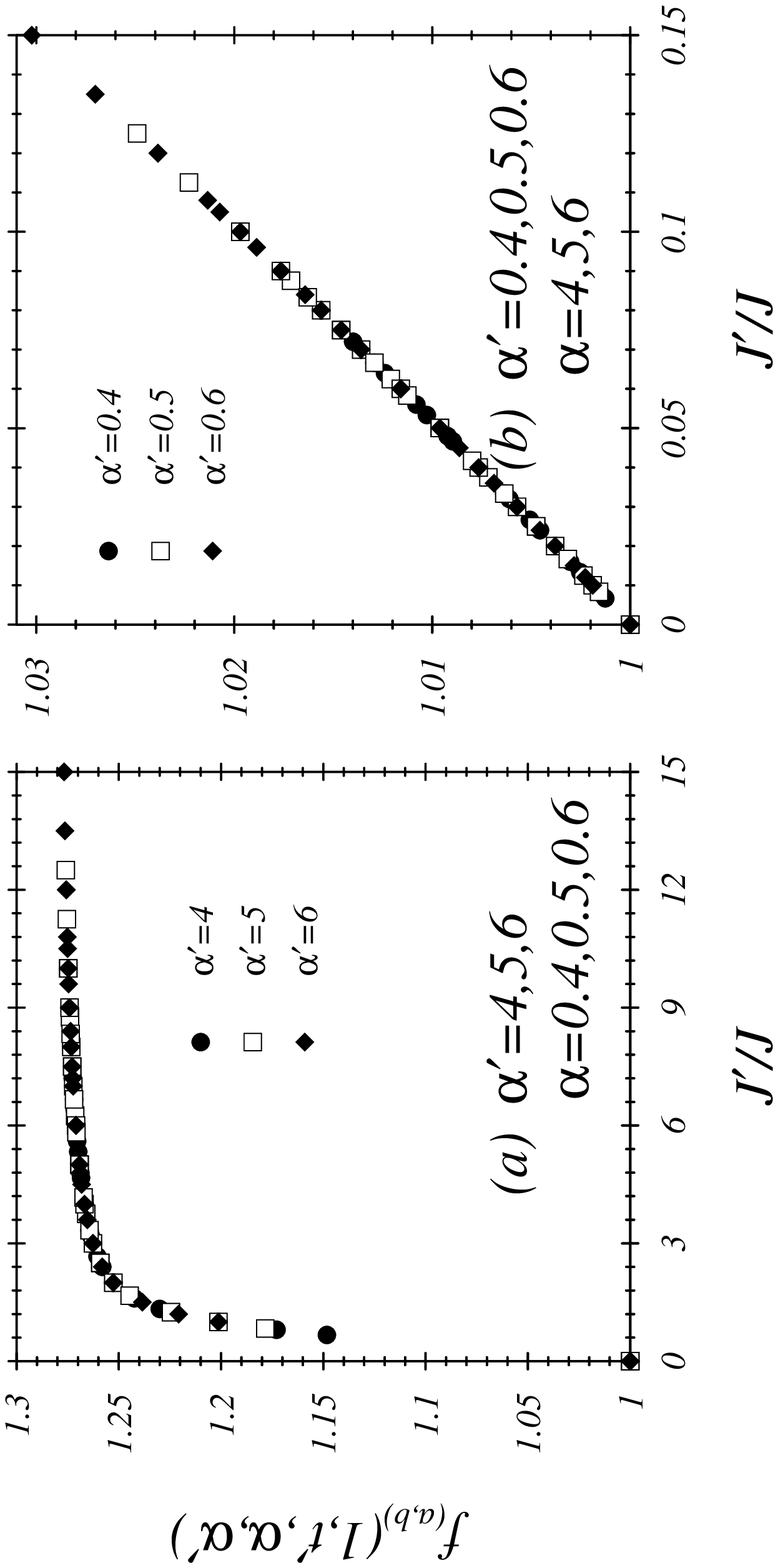}}
\caption{Correction factor $f(1,t',\alpha ,\alpha')$ of
(\ref{frac_eps_2}) for a $2\times 8$ $t-J$ ladder with
parameters (\ref{coupl_parm_2})(a),(b) and $\rho=1$}
\label{nfig3}
\end{figure}

Let us next discuss the linear $\rho$-behaviour in (\ref{eps0_gt}).
The dependence on the parameters $t',\alpha,\alpha'$ enters via
$\varepsilon(1/2,t',\alpha,\alpha')$ [(\ref{frac_eps}),(\ref{Ep})(a),(\ref{Ep})(b)]
and $\varepsilon(1,J=\alpha,J'=\alpha't')$. 
Note that for $\rho=1$,
$\varepsilon(1,J,J')$ is just the ground state energy per site of
the spin 1/2 Heisenberg model on a two leg ladder with spin couplings
$J$ and $J'$ along the rungs and legs, respectively.\footnote{Note that
at $\rho=1$ the
Heisenberg and $t-J$ Hamiltonians are differing by a constant diagonal
contribution ($-1/4$) per bond.} The $J,J'$ dependence
\begin{eqnarray}
\varepsilon(1,J,J') & = &
\frac{1}{4}E^{(p)}(4,J,J')\cdot f(1,J'/J)
\nonumber\\
 & & \label{frac_eps_2}
\end{eqnarray}
is correctly reproduced by the plaquette ground state energy $E^{(p)}(Q,J,J')$
with charge $Q=4$ [cf. (\ref{e04_s})]:
\begin{eqnarray}\label{E_Q4}
E^{(p)}(Q=4,J,J') & = & -J-J'-\sqrt{J^2+J'^2-JJ'}\nonumber\\
 & & \label{Ep_Q4}
\end{eqnarray}
up to a correction factor $f(1,J'/J)$, which is shown in Figs. \ref{nfig3}(a),
\ref{nfig3}(b) and which scales in $J'/J$.

\section{Plaquette clusters in the ground state of the $t-J$
model on a two leg ladder\label{sec3}}

The results of the numerical analysis in the last section
motivate us to build up the ground state on the two leg ladder
from plaquette eigenstates.

\begin{figure}[ht!]
\centerline{\includegraphics[width=8.0cm,angle=0]{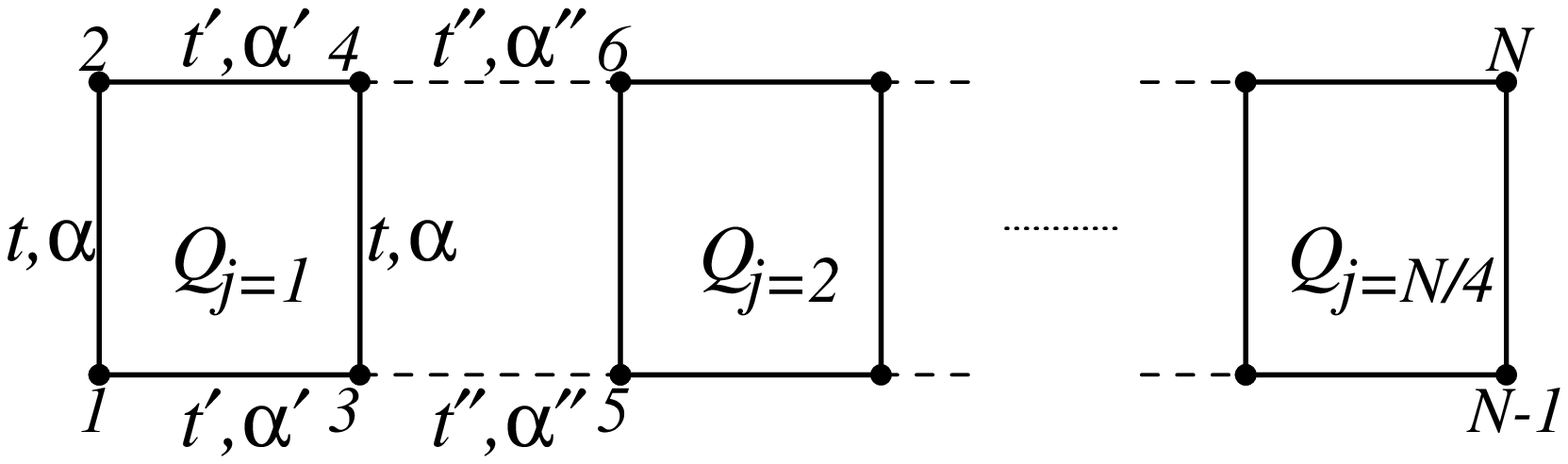}}
\caption{Decomposition of the two leg $t-J$ ladder into coupled
($t'',\alpha''$) 4-site plaquette clusters}
\label{fig1}
\end{figure}

For this purpose the $t-J$ Hamiltonian
\begin{eqnarray}
H & = & t\sum_{j=1}^{N/4}h_{j,j}(t',\alpha',\alpha)+t''\sum_{j=1}^{N/4-1}
h_{j,j+1}(\alpha'')\nonumber\\
\label{H_tJ}
\end{eqnarray}


is decomposed
into $N/4$ plaquette Hamiltonians [using the notation of
Ref. (\onlinecite{fl03})]
\begin{eqnarray}\label{h_jj}
h_{j,j}(t',\alpha',\alpha)=\hspace{5.0cm} &  & \nonumber\\
\hspace{0.5cm}\frac{t'}{t}\left[h(4j-3,4j-1,\alpha')+
h(4j-2,4j,\alpha')\right]+ & &\nonumber\\
\hspace{-0.0cm}\left[h(4j-3,4j-2,\alpha)+h(4j-1,4j,\alpha)\right]\hspace{0.5cm} & &
\end{eqnarray}
with spin couplings $J=t\alpha$, $J'=t'\alpha'$ and hopping terms
$t,t'$ along the rungs and legs, respectively (cf. Fig. \ref{fig1}).
\begin{eqnarray}\label{h_j_j+1}
h_{j,j+1}(\alpha'')=\hspace{5.0cm} &  & \nonumber\\
\left[h(4j-1,4j+1,\alpha'')+h(4j,4j+2,\alpha'')\right]\hspace{0.5cm} & &
\end{eqnarray}
describes the interaction between neighbouring plaquettes $j,j+1$.
This interaction is treated in the following in a perturbative expansion
around $t''=0$, $J''=0$, $\alpha'=\alpha''=J''/t''$ fixed.
In the analysis of the numerical results for the ground state energies
(\ref{frac_eps}) and (\ref{frac_eps_2}) at $\rho=1/2$ and $\rho=1$,
this interaction generates the correction factors $f(\rho=1/2)$ and
$f(\rho=1)$ -- shown in Figs. \ref{nfig2}(a),(b) and \ref{nfig3}(a),(b)
respectively.

The lowest plaquette eigenstates
\begin{eqnarray}\label{SE}
h_{j,j}(t',\alpha',\alpha)\psi_{n_j}^{(p)}(Q_j) & = & E_{n_j}^{(p)}(Q_j)
\psi_{n_j}^{(p)}(Q_j)
\end{eqnarray}
in the sector with charge $Q_j=0,2,4$, $Q_j=1,3$ and total plaquette spin 0
and $1/2$, respectively,
are discussed in Appendix \ref{appendix_a}. The eigenstates
$\psi_{n_j=0}^{(p)}(Q_j)=\psi^{(p)}(Q_j)$ with lowest energy 
$E_{n_j=0}^{(p)}(Q_j)=E^{(p)}(Q_j)$
yield the basis for the ground state on the two leg ladder in
lowest order perturbation theory $t''=0$.

To zeroth order in $t''$ ($t''=0$) the eigenstates of the Hamiltonian
(\ref{H_tJ}) are given by a product of plaquette ground states
$\psi^{(p)}(Q_j)$
\begin{eqnarray}\label{plaq_prod}
 & & \prod_{j=1}^{N/4}\psi^{(p)}(Q_j)
\end{eqnarray}
with energies
\begin{eqnarray}
E_n & = & \sum_{j=1}^{N/4}E^{(p)}(Q_j)\,.
\end{eqnarray}
The plaquette charges $Q_j$ have to add up to the total charge:
\begin{eqnarray}
Q_{\mbox{tot}} & = & \sum_jQ_j\,.
\end{eqnarray}
Our analysis in Sec. \ref{sec2} suggests that only plaquette
ground states with charges $Q=0,2,4$:
\begin{eqnarray}\label{psi_Q2}
\hspace{-0.8cm}\psi^{(p)}(0),\,\psi^{(p)}(2) & &
\mbox{for}\quad\quad 0\leq\rho\leq 1/2\\
\hspace{-0.8cm}\psi^{(p)}(2),\,\psi^{(p)}(4) & &
\mbox{for}\quad 1/2\leq\rho\leq 1\label{psi_Q4}
\end{eqnarray}
are involved in the construction of the ground state of the two
leg ladder.
The corresponding ground state energies on the ladder:
\begin{eqnarray}\label{e0_cl_prod}
E(\rho,t',\alpha',\alpha,t''=0) & = &
N^{(i)}(i)E_0^{(p)}(Q=i)\\
 & & +N^{(i)}(i+2)E_0^{(p)}(Q=i+2)\,.\nonumber\\
i=0\quad\mbox{for} & & \quad 0\leq\rho\leq 1/2\\
i=2\quad\mbox{for} & & 1/2\leq\rho\leq 1
\end{eqnarray}
are obtained from plaquette ground state energies $E^{(p)}(Q)$
and the number $N^{(i)}(Q)$ of plaquettes with charge $Q$
\begin{eqnarray}
N^{(0)}(0)+N^{(0)}(2) & = & N^{(2)}(2)+N^{(2)}(4)=N/4\quad\quad\quad\quad\\
2N^{(0)}(2) & = & 2N^{(2)}(2)+4N^{(2)}(4)=Q_{\mbox{tot}}\,.
\end{eqnarray}
So far we only treated the zeroth order perturbation theory
(\ref{H_tJ}) ($t''=0$). The product states are degenerate
since the $N^{(i)}(Q)$ plaquettes with charges $Q=0$, $Q=2$
for $i=0$ and $Q=2,Q=4$ for $i=2$ can be distributed in different
ways over the ladder (Fig. \ref{fig1}).

The interaction energy $W(Q_j,Q_{j+1})$ between neighbouring
plaquettes is derived in Appendix \ref{appendix_b} in the
framework of a $1^{st}$ order perturbation theory in the
hopping parameter $t''$. The resulting shift $\Delta E$ in 
the ground state energy:
\begin{eqnarray}
\Delta E^{(0)} & = & N^{(0)}(2,2)W(2,2)\label{D0}\\
 & & \quad\mbox{for}\,\,\,i=0,\quad 0\leq\rho\leq 1/2\,;\nonumber\\[4pt]
\Delta E^{(2)} & = & N^{(2)}(2,2)W(2,2)+N^{(2)}(2,4)W(2,4)\nonumber\\
 & & +N^{(2)}(4,4)W(4,4)\label{D2}\\
 & & \quad\mbox{for}\,\,\,i=2,\quad 1/2\leq\rho\leq 1\,.\nonumber
\end{eqnarray}
can be expressed in terms of the interaction energies $W(Q_j,Q_{j+1})$
and numbers $N(Q_j,Q_{j+1})$ of neighbouring plaquettes with charges
$Q_j,Q_{j+1}$. 
Since $W(0,0)=W(0,2)=0$ and $W(2,2)=-J'/8$ (\ref{W_Qj_Qj+1}), 
$\Delta E^{(0)}$ is minimal if $N^{(0)}(2,2)$ is maximal:
\begin{eqnarray}
N^{(0)}(2,2) & = & N^{(0)}(2)-1=
\frac{N}{2}\rho-1\,,
\end{eqnarray}
According to (\ref{W_Qj_Qj+1}):
\begin{eqnarray}
W(2,2)+W(4,4)-2W(2,4) & < & 0\,.
\end{eqnarray}
$\Delta E^{(2)}$ is minimal if $N^{(2)}(4,4)$ (and thereby
$N^{(2)}(2,2)$) are maximal.
\begin{eqnarray}
N^{(2)}(4,4) & = & N^{(2)}(4)-1=\frac{N}{2}(\rho-1/2)-1\,,\quad\quad\\
N^{(2)}(2,2) & = & N^{(2)}(2)-1=\frac{N}{2}(1-\rho)-1\,,\\
N^{(2)}(2,4) & = & 1\,,
\end{eqnarray}
The perturbative result of the ground state energy per site in
the regime (\ref{coupl_parm_2})(a)
\begin{eqnarray}
\hspace{-0.8cm}\varepsilon(\rho) & = & (1-\rho)\frac{1}{2}\Big(E^{(p)}(2)+
W(2,2)\Big)\nonumber\\
 & &
 +\frac{1}{2}(\rho-1/2)\Big(E^{(p)}(4)
 +W(4,4)\Big)\label{e0_pert_2}
\end{eqnarray}
predicts a level crossing in all charge sectors with $1/2<\rho<1$ if
\begin{eqnarray}
E^{(p)}(4)-E^{(p)}(2)+W(4,4)-W(2,2) & = & 0\,.\nonumber\\
 & & \label{crossing}
\end{eqnarray}

In the regime (\ref{coupl_parm_2})(a) the left-hand side with
(\ref{Ep_Q4}), (\ref{Ep})(a) and
\begin{eqnarray}
W(4,4)-W(2,2)
& = & -\frac{3}{8}J'
\end{eqnarray}
only depends on $J,J'$. The solution defines a curve $J'=J'(J)$ in the
parameter space, where the ground state energy (\ref{e0_pert_2})
becomes independent of $\rho$, the corresponding chemical potential:
\begin{eqnarray}
\mu=\frac{d\varepsilon}{d\rho} & = & 0\quad\mbox{for}\quad
1/2\leq\rho\leq 1\quad \mbox{and}\quad J'=J'(J)\nonumber\\
 & &
\end{eqnarray}
turns out to be zero here.

Moreover, the ground state energy along the curve $J'(J)$ 
is predicted to be
\begin{eqnarray}\label{eps_jp}
\varepsilon(\rho,J'=J'(J)) & = & \frac{1}{4}E_0^{(p)}(2)-\frac{J'}{32}\,.
\end{eqnarray}
Results for $J'(J)$ and $\varepsilon(\rho,J'=J'(J))$ are given
in Table \ref{table_e_JJp}.
\begin{table}[ht!]
\begin{tabular}{|c||r|r|r|}
  & & & \\[-13pt] \hline
 $J=\alpha$ & $0.4$\quad\quad\quad   &  $0.5$\quad\quad\quad   &  
$0.6$\quad\quad\quad \\ \hline
 & & & \\[-12pt] \hline
 $J'=J'(J)$ & \quad 0.94895\quad\quad  & \quad 0.88839\quad\quad &
\quad 0.81255\quad\quad \\ \hline
 $\varepsilon(\rho,J'(J))$ & \quad-0.66215\quad\quad   & 
\quad-0.65099\quad\quad  & 
\quad-0.63717\quad\quad\\[2pt]\hline
\end{tabular}
\caption{Results for $J'(J)$ and $\varepsilon(\rho,J'=J'(J))$ for
rung spin exchange couplings $J=\alpha=0.4,0.5,0.6$ on the basis
of (\ref{crossing}) and (\ref{eps_jp}).}
\label{table_e_JJp}
\end{table}

\begin{figure}[ht!]
\centerline{\includegraphics[height=8.0cm,angle=-90]{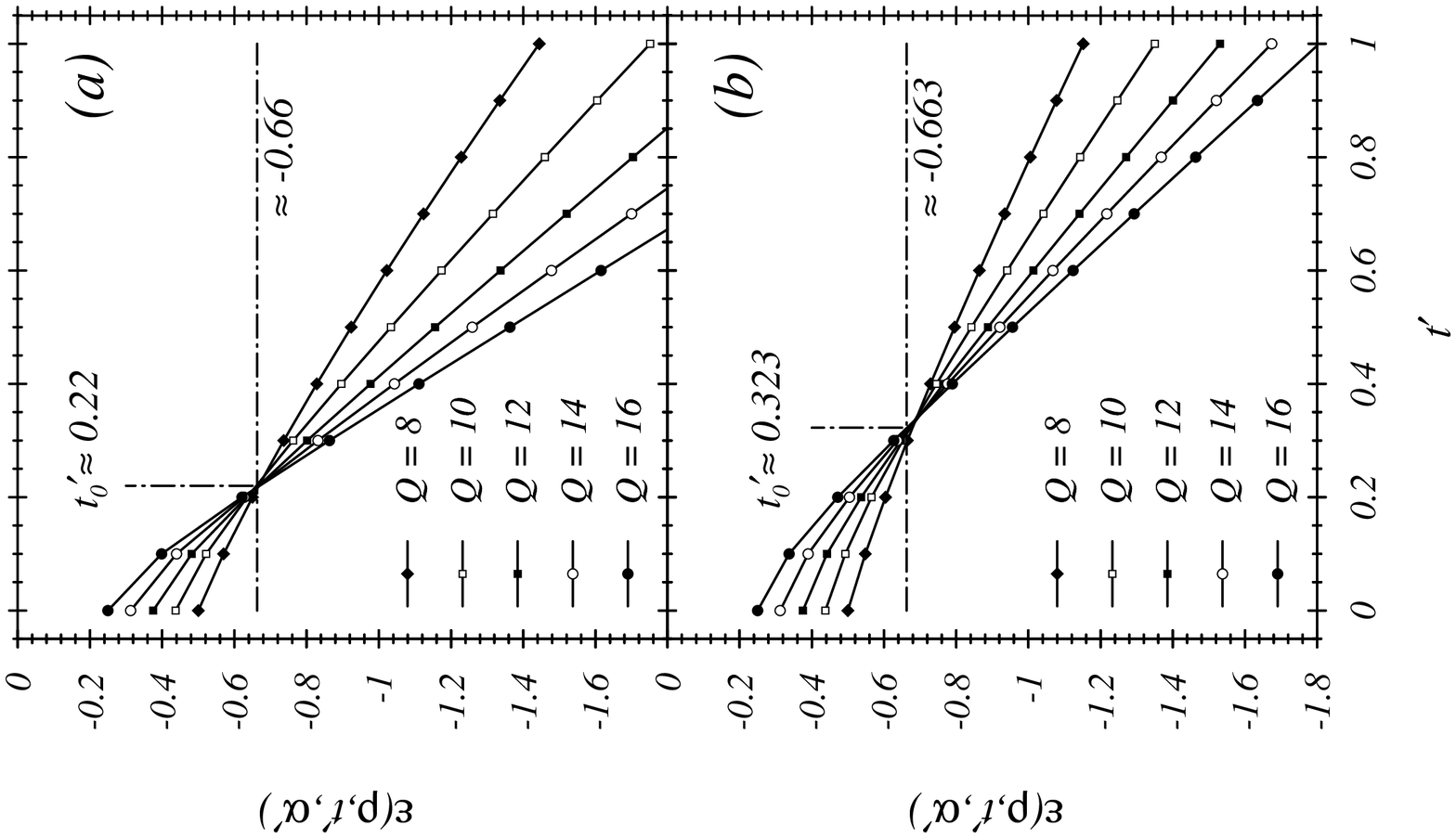}}
\caption{Lanczos results for the crossing of energy levels for 
$\rho=Q/N\ge 1/2$ and $\alpha=0.5$, 
$\alpha'=4.0(a),\alpha'=2.7(b)$ for a $N=2\times 8$ $t-J$ ladder}
\label{fig3}
\end{figure}


As an illustration, we present in Fig. \ref{fig3} the ground state energy
per site $\varepsilon(\rho,t',\alpha')=E(N,Q,t',\alpha',\alpha=1/2)/N$ for the charges
$Q=8,10,\ldots,16$ at $\alpha=0.5$, $\alpha'=4.0$ (a) and
$\alpha'=2.7$ (b). 

The crossing of these energy levels at
\begin{eqnarray}
t'_0 & = & \frac{J'(J=0.5)}{\alpha'}=\left\{
\begin{array}{lcc}0.222 & \mbox{for} & \alpha'=4.0\\
0.329 &  \mbox{for} & \alpha'=2.7 \end{array}\right.\nonumber\\
 & & \label{t0p_alp}
\end{eqnarray}
is predicted to change with $\alpha'$ if we keep $\alpha=J=0.5$ fixed.
On the other hand, the corresponding ground state energy per site
\begin{eqnarray}
\varepsilon(\rho,J'(J=0.5)) & = & -0.65099
\label{eps_t0p_alp}
\end{eqnarray}
is independent of $\alpha'$! 

Both predictions (\ref{t0p_alp}) and 
(\ref{eps_t0p_alp})
are clearly visible in the numerical results on a 
$2\times 8$ ladder.



We also looked for level crossings (\ref{crossing}) in the
regime (\ref{coupl_parm_2})(b). It turns out that they occur
at $t'$ values $t''=t'>1$, where the perturbative approach
is not reliable.

\section{The special role of plaquette clusters with charges
$Q=2$ and $Q=4$\label{sec4}}

The variational ansatz (\ref{plaq_prod}) with plaquette clusters
only involves cluster eigenstates (\ref{psi_Q2}), (\ref{psi_Q4})
with even charges $Q=0,2,4$. Such an ansatz makes sense if
plaquette clusters with odd charges are suppressed energetically:
\begin{eqnarray}
\Delta_1 & = & E^{(p)}(0)+E^{(p)}(2)-2E^{(p)}(1)<0\label{ineq_0211}\,,\\
\Delta_3 & = & E^{(p)}(2)+E^{(p)}(4)-2E^{(p)}(3)<0\label{ineq_2433}\,.
\end{eqnarray}
In Appendix \ref{appendix_a} we discuss the low-lying eigenstates
of plaquette clusters with charges $Q=0,1,2,3,4$. The ground state
energies for $Q=0,1,4$ are unique in the sense that there are
no level crossings by variation of the parameters $t',\alpha',\alpha$.
The ground state energy of the $Q=2$ cluster is given by
(\ref{Ep})(a) and (\ref{Ep})(b), respectively.

It turns out that the inequalities (\ref{ineq_0211}) and
(\ref{ineq_2433}) are indeed satisfied in the regimes
(\ref{tuning})(a) and (\ref{tuning})(b), respectively.

%
%



The suppression of $Q=3$ plaquettes in the regime (\ref{ineq_2433})
has immediate consequences for the mobility and correlations of holes
as they are discussed by Siller, Troyer, Rice and White\cite{siller01}
for the low doping case ($\rho>3/4$). 
In the regime $\Delta_3<0$ hole pairs are confined in $Q=2$ plaquettes
and cannot move in the antiferromagnetic background of the $Q=4$
plaquettes. This can be seen in the first order perturbation
theory (cf. Appendix \ref{appendix_b}) for the interaction between
neighbouring plaquettes, which forbids the transition
\begin{eqnarray}
Q_j=2\quad Q_{j+1}=4 & \rightarrow & Q_j'=4\quad Q_{j+1}'=2
\label{Q_24_42}
\end{eqnarray}
with charge transfer $\Delta Q=2$. Therefore the hopping of the
hole pair confined in the $Q=2$ plaquette is suppressed in the
antiferromagnetic background.

On the other hand hopping of a single hole contained in a plaquette
with $Q=3$ is possible since the transition
\begin{eqnarray}
Q_j=3\quad Q_{j+1}=4 & \rightarrow & Q_j'=4\quad Q_{j+1}'=3
\label{1st_allowed}
\end{eqnarray}
is not forbidden.

It is interesting to study the energy difference $\Delta_3(\alpha,\alpha')$
for the $\alpha,\alpha'$ values of Ref. (\onlinecite{siller01}). There
the leg coupling $\alpha'=J'/t'$ has been chosen to $\alpha'=0.35$ 
whereas the rung coupling varies $\alpha\geq\alpha'=0.35$. The hopping
parameters are equal $t=t'$. For large $\alpha>4$ this corresponds to
our regime (\ref{coupl_parm_2})(b).

\begin{figure}[ht!]
\centerline{\includegraphics[height=8.0cm,angle=-90]{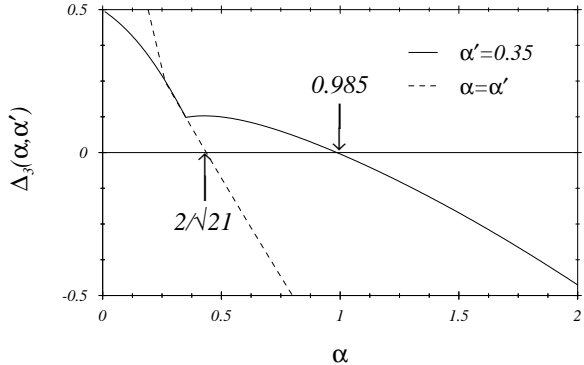}}
\caption{ Energy differences $\Delta_3(\alpha,\alpha')$ for (1) $\alpha'=0.35$
and (2) $\alpha=\alpha'$ (in both cases: $t=t'=1$).}
\label{fig5vf}
\end{figure}

In Fig. \ref{fig5vf} we have plotted the difference
$\Delta_3(\alpha,\alpha'=0.35)$. For comparison we have also included the
difference $\Delta_3(\alpha=\alpha')$ with symmetric couplings along
the rungs and legs, respectively.
The latter drops monotonically with $\alpha=\alpha'$ and has zero at
\begin{eqnarray}
\alpha=\alpha' & = & \frac{2}{\sqrt{21}}=0.436...
\label{zero_symm}
\end{eqnarray}

In contrast the difference $\Delta_3(\alpha,\alpha'=0.35)$ in the
asymmetric case $\alpha>\alpha'=0.35$ first increases and has a flat
maximum at $\alpha\simeq 0.43$ and then drops with a zero at
\begin{eqnarray}
\Delta_3(\alpha\simeq 0.985,\alpha'=0.35) & = & 0\,.
\label{zero_0.35}
\end{eqnarray}
The zeros (\ref{zero_symm}) and (\ref{zero_0.35}) of $\Delta_3$ mark
the transition where $Q=3$ plaquettes are substituted by $Q=2$ plaquettes
and 2 holes combine to a pair. We expect that the hole-hole correlation
length has a maximum at this transition point. 
Indeed there is a maximum of the hole-hole correlation length at
$\alpha=1.2$, as determined in Ref. (\onlinecite{siller01}) from
a DMRG calculation on a $40\times 2$ ladder. It might be an accident,
that the value $\alpha=1.2$ is quoted as well as a lower bound for phase 
separation in the $2D$ $t-J$ model with isotropic couplings (Putikka et
al.\cite{putikka92}). We therefore calculated (\ref{ineq_2433}) as well
for the isotropic case $\alpha=\alpha'$, where the zero (\ref{zero_symm})
is formed quite below $\alpha=1.2$.

It would be interesting to see whether
the maximum of the correlation length is shifted as well to a smaller
value in the symmetric case $\alpha=\alpha'$.

We are aware of the fact, that our considerations in the low doping
regime ($\rho>3/4$) are based on the product ansatz (\ref{plaq_prod})
with plaquette clusters with charges
\begin{eqnarray}
Q=2,4 & \mbox{for\quad} & \Delta_3<0\label{D3_lt_0}\\
Q=3,4 & \mbox{for\quad} & \Delta_3>0\label{D3_gt_0}
\end{eqnarray}
The interaction between neighbouring plaquettes is neglected.
This is justified for (\ref{D3_lt_0}), if $\Delta_3$ is sufficiently
negative, as was demonstrated in Secs. \ref{sec2} and \ref{sec3}.
On the other hand, these interactions cannot be neglected in the
regime (\ref{D3_gt_0}), where first order perturbation theory
allows the hopping (\ref{1st_allowed}) of $Q=3$ plaquettes in the
antiferromagnetic background of $Q=4$ plaquettes. In this case,
first order perturbation theory leads to an effective Hamiltonian
on a chain with nearest neighbour couplings. The effective degrees
of freedom at each site and their nearest neighbour interactions
are defined by the ground states of the $Q=3$ and $Q=4$ plaquettes.
We expect, that such an effective Hamiltonian will induce a convex
curvature in the $\rho$-dependence of the ground state energy, which
would indicate that we are beyond phase separated phase. We observed
this curvature (for $0<\rho<1/2$) with an effective Hamiltonian
based on rung clusters and their interactions.\cite{fl03}


\section{Discussion and perspectives\label{sec5}}

The formation of clusters in the ground state of a quasi 
one-dimensional system has important consequences for its 
physical properties, e.g. the phase diagram at zero
temperature.

In case of the $t-J$ model on a two leg ladder with
asymmetric couplings ($\alpha=J/t,\alpha'=J'/t'$;
regime (\ref{coupl_parm_2})(a) $\alpha/\alpha'\ll 1$ and
regime (\ref{coupl_parm_2})(b) $\alpha/\alpha'\gg 1$)
plaquette clusters with even charges $Q=0,2,4$ play the
dominant role and explain the charge density and
($\alpha,\alpha',t'$) dependence of the ground state energies,
as was demonstrated in Secs. \ref{sec3} and \ref{sec4}.
First order perturbation theory for  the interaction of
neighbouring plaquettes favour the condensation of plaquettes
with the same charge ($Q=2,Q=4$), which can be interpreted as
a signal for phase separation. Of course, this can happen only
if plaquette clusters with odd charges $Q=1,Q=3$ are suppressed 
energetically, which means that the energy combinations
(\ref{ineq_0211}) (\ref{ineq_2433}) are sufficiently negative.

The low doping regime ($\rho\geq 3/4$) is of special interest.
It has been demonstrated in Ref. (\onlinecite{siller01}) that
the hole-hole correlations are small in the $t-J$ model with
asymmetric couplings on legs ($\alpha'=0.35$) and rungs
($\alpha>4$). In this regime the ground state is very well
described with plaquette clusters of charge $Q=2$, $Q=4$.
The $Q=4$ plaquettes generate the antiferromagnetic background.
In each $Q=2$ plaquette a pair of holes is confined. The holes
can be deconfined only if the resulting two plaquettes with
charge $Q=3$ are energetically preferred [cf. (\ref{ineq_2433}) for
$\Delta_3>0$]. As was demonstrated in Sec. \ref{sec4} this happens
for smaller $\alpha$ values ($\alpha'=0.35<\alpha<0.98$ in the
asymmetric, $\alpha=\alpha'<2/\sqrt{21}$ in the symmetric case).
In this regime the ground state is more complex [Ref. (\onlinecite{siller01})].

In our approach based on a product ansatz with clusters we have
two possibilities to exploit the ground state in this regime:
(A) We compute the effective Hamiltonian which describes the
interaction between $Q=3$ and $Q=4$ plaquettes perturbatively.
(B) We improve the quality of the product ansatz with larger
clusters such that hole-hole correlations at larger distances
are properly taken into account as well.

\acknowledgments

We are indebted to A. Kl\"umper for a critical
reading of the manuscript. 

 
\begin{appendix}
\section{Eigenstates of the 4-site cluster}
\label{appendix_a}          

\begin{figure}[ht!]
\centerline{\includegraphics[width=6.0cm,angle=0]{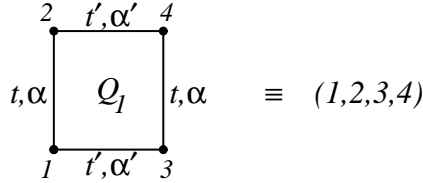}}
\caption{Notation and shape of the used 4-site cluster -- building
block for the two leg $t-J$ ladder}
\label{figa1}
\end{figure}

The Hamiltonian of the 4-site cluster (cf. Fig. \ref{figa1})
is defined in (\ref{H_tJ}) (for $j=1$).

We are now going to construct all $N_c(Q)$ eigenstates for
cluster charges $Q=Q_1=0,1,2,3,4$ -- assuming equal numbers of spin-up
($+$) and spin-down ($-$) charges for even $Q$ and an excess of one
spin-up particle for odd cluster charges $Q$. Moreover, we will
assume $t,J,t',J'\ge 0$.

The difference in coupling parameters for rungs ($t,J$) and
legs ($t',J'$) subdivides the basis states for each $Q$ into
the classes $\{|III\rangle\}$ and $\{|I\rangle,|II\rangle\}$

\begin{eqnarray}
 & & Q=1\quad\quad\quad Q=2\quad\quad\quad Q=3\quad\quad Q=4\nonumber\\[5pt]
|III\rangle : &  & \left|\begin{array}{cc}
0 & 0\\ + & 0\end{array}\right\rangle,\quad
\left|\begin{array}{cc}0 & -\\ + & 0\end{array}\right\rangle,\quad
\left|\begin{array}{cc}+ & 0\\ - & +\end{array}\right\rangle,\quad
\left|\begin{array}{cc}- & +\\ + & -\end{array}\right\rangle,
\nonumber\\
 &  & \left|\begin{array}{cc}
+ & 0\\ 0 & 0\end{array}\right\rangle,\quad
\left|\begin{array}{cc}+ & 0\\ 0 & -\end{array}\right\rangle,\quad
\left|\begin{array}{cc}- & +\\ + & 0\end{array}\right\rangle,\quad
\left|\begin{array}{cc}+ & -\\ - & +\end{array}\right\rangle,
\nonumber\\
 &  &  \label{states_3}
\end{eqnarray}
and\footnote{Note, that each shown state represents 2 elements
due to successive $\pi$-rotations -- except the $Q=4$ states
in (\ref{states_3}).}
\begin{eqnarray}
|I\rangle : &  &
\hspace{2.0cm}
\left|\begin{array}{cc}- & 0\\ + & 0\end{array}\right\rangle,\quad
\left|\begin{array}{cc}- & 0\\ + & +\end{array}\right\rangle,\quad
\left|\begin{array}{cc}- & -\\ + & +\end{array}\right\rangle,\nonumber\\
 & & \hspace{2.0cm}
\left|\begin{array}{cc}+ & 0\\ - & 0\end{array}\right\rangle,\quad
\left|\begin{array}{cc}+ & +\\ - & 0\end{array}\right\rangle,\quad\nonumber\\
 &  &  \label{states_1}
\end{eqnarray}
\begin{eqnarray}
|II\rangle : &  &
\hspace{2.0cm}
\left|\begin{array}{cc}0 & 0\\ + & -\end{array}\right\rangle,\quad
\left|\begin{array}{cc}+ & 0\\ + & -\end{array}\right\rangle,\quad 
\left|\begin{array}{cc}+ & -\\ + & -\end{array}\right\rangle,\nonumber\\
 & & \hspace{2.0cm}
\left|\begin{array}{cc}0 & 0\\ - & +\end{array}\right\rangle,\quad
\left|\begin{array}{cc}+ & -\\ + & 0\end{array}\right\rangle,\quad\nonumber\\
 &  &  \label{states_2}
\end{eqnarray}
concerning the behaviour with respect to transformations
\begin{eqnarray}
\hat R:\quad(1,2,3,4) & \longrightarrow & (1,3,2,4)\,,
\end{eqnarray}
i.e. the interchange of legs and rungs.

\begin{itemize}
\item[a)]
$Q=0$, \,\,$N_c(0)=1$

\begin{eqnarray}
\hspace{0.6cm}E^{(p)}(Q=0) & =  & 0\,.
\end{eqnarray}

\item[b)]
$Q=1$, \,\,$N_c(1)=4$

We introduce $|n\rangle$ as $Q=1$-basis state with the spin-up
particle ($+$) at plaquette position $n$ ($|n\rangle\equiv c_{n\uparrow}^+|0\rangle$).

Using the basis:
\begin{eqnarray}
\hspace{0.0cm}|1,\tau,\zeta\rangle & = & \frac{1}{2}
(|1\rangle + \tau|4\rangle)+\zeta
(|2\rangle + \tau|3\rangle)
\end{eqnarray}
with $\tau=\pm 1$, $\zeta=\pm 1$, we obtain the eigenvalues:
\begin{eqnarray}
\hspace{0.6cm}E^{(p)}(Q=1,\tau,\zeta) & =  & 
-\zeta(t+\tau t')\,.
\end{eqnarray}

\item[c)]
$Q=2$, \,\,$N_c(2)=12$

Introducing $|m,n\rangle$ as $Q=2$-basis state with spin-up
particles ($+$) at site $m$ and ($-$) at position
$n$ ($|m,n\rangle\equiv c_{m\uparrow}^+c_{n\downarrow}^+|0\rangle$), 
we first compute the action of the $t-J$ Hamiltonian
onto the singlet ($\tau=1$) and triplet states ($\tau=-1$):
\begin{eqnarray}
|1,\tau,\zeta\rangle & = & |1,2\rangle +\tau|2,1\rangle+\zeta\left(|3,4\rangle
+\tau|4,3\rangle\right)\nonumber\\
|2,\tau,\zeta\rangle & = & |2,4\rangle +\tau|4,2\rangle+\zeta\left(|1,3\rangle
+\tau|3,1\rangle\right)\label{Q2_states}\\
|3,\tau,\zeta\rangle & = & |1,4\rangle +\tau|4,1\rangle+\zeta\left(|3,2\rangle
+\tau|2,3\rangle\right)\nonumber\,.
\end{eqnarray}
In this basis the action of the $t-J$ Hamiltonian results in
a $3\times 3$ matrix for the singlet ($\tau=1$) sector
\begin{eqnarray}
 & & \left(\begin{array}{ccc}-J & 0 & -(1+\zeta)t'\\ 0 & -J' & -(1+\zeta)t\\
-(1+\zeta)t' & -(1+\zeta)t & 0\end{array}\right)\,.
\label{Q2_3x3}
\end{eqnarray}
The ground state energy $E$ in the singlet sector ($\tau=1$ with $\zeta=1$)
is found from the solution of the third order equation
\begin{eqnarray}
\hspace{1.0cm}-(J+E)(J'+E)E+4(E+J)t^2+4(E+J')t'^2 &  & \nonumber\\
 \hspace{0.0cm}=0\hspace{1.7cm} & & \label{3rd_order}
\end{eqnarray}
which can be easily solved in the symmetric case $t'=t$, $J'=J$:
\begin{eqnarray}
E & = & -\frac{t}{2}\left(J+\sqrt{J^2+32}\right)\,.
\end{eqnarray}
In the asymmetric case (\ref{coupl_parm_2})(a) with $\alpha=J/t\ll
\alpha'=J'/t'$ one can derive an iterative solution treating the
term
\begin{eqnarray}
-(J'+E)E+4t^2 & = & -4\frac{E+J'}{E+J}t'^2\label{itera_sol}
\end{eqnarray}
on the right-hand side of (\ref{itera_sol}) as a perturbation.
The resulting ground state energy in first order of this
perturbation reads:
\begin{eqnarray}
E_0^{(p)} & = & E_a^{(p)}+\Delta E_a\label{E0_p}\\
\Delta E_a & = & -4\frac{E_a^{(p)}+J'}{E_a^{(p)}+J}\frac{t'^2}{\sqrt{J'^2+16t^2}}
\label{Delta_Ea}
\end{eqnarray}
where
\begin{eqnarray}
E_a^{(p)} & = & -\frac{1}{2}\left(J'+\sqrt{J'^2+16t^2}\right)\,.
\label{E_p_a}
\end{eqnarray}
E.g. for $J=0.5$, $J'=4$, $t'=1$ the correction term (\ref{Delta_Ea})
yields a 3\% contribution to the ground state energy $E$, such that
the zeroth order $E_a^{(p)}$ is already a very good approximation.

In the regime (\ref{coupl_parm_2})(b) with $\alpha'=J'/t'\ll\alpha=J/t$
we derive from (\ref{3rd_order}) a corresponding ground state energy in a 
first order perturbation:
\begin{eqnarray}
E_0^{(p)} & = & E_b^{(p)}+\Delta E_b\\
\Delta E_b & = & -4\frac{E_b+J}{E_b+J'}\frac{t'^2}{\sqrt{J^2+16t'^2}}
\label{Delta_Eb}
\end{eqnarray}
where
\begin{eqnarray}
E_b^{(p)} & = & -\frac{1}{2}\left(J+\sqrt{J^2+16t'^2}\right)\,.
\label{E_p_b}
\end{eqnarray}
Let us now turn to the triplet sectors ($\tau=-1$).
Here the 6 eigenstates and corresponding eigenvalues turn out to be
\begin{eqnarray}
|1,(-1,-1)\rangle & & E=0\nonumber\\
|2,(-1,-1)\rangle & & E=0\nonumber\\
|1,(-1,1)\rangle\pm
|3,(-1,1)\rangle & &  E=\pm 2t'\nonumber\\
|2,(-1,1)\rangle\pm
|3,(-1,-1)\rangle & & E=\pm 2t\,.\nonumber
\end{eqnarray}

\item[d)]
$Q=3$, \,\,$N_c(3)=12$

We introduce $|m,n\rangle$ as $Q=3$-basis state with the hole
($0$) at plaquette position $m$ and spin-down electron ($-$) at position
$n$.
The creation operators of the
two spin-up electrons $(+)$ and that of the spin-down electron
$(-)$ are ordered according to the increasing site number (cf. Fig. \ref{figa1})A.
E.g. $|1,3 \rangle\equiv c_{2\uparrow}^+
c_{3\downarrow}^+c_{4\uparrow}^+|0\rangle$.

In the basis:
\begin{eqnarray}
\hspace{0.6cm}|1,\tau,\zeta\rangle & = & (|4,2\rangle + \tau|1,3\rangle)
+\zeta (|3,1\rangle + \tau|2,4\rangle)\nonumber\\
|2,\tau,\zeta\rangle & = & (|4,3\rangle + \tau|1,2\rangle)
+\zeta (|3,4\rangle + \tau|2,1\rangle)\nonumber\\
|3,\tau,\zeta\rangle & = & (|4,1\rangle + \tau|1,4\rangle)
+\zeta (|3,2\rangle + \tau|2,3\rangle)\nonumber\\
\end{eqnarray}
the $t-J$ Hamiltonian reduces to the following $3\times 3$ matrix
for the eigenvalues with $x=J/2$, $x'=J'/2$:
\begin{eqnarray}
 & & \left(\begin{array}{ccc} \zeta\tau t'-x & 0 & -\zeta t+x\\ 
0 & -\zeta t-x' & \zeta\tau t'+x' \\
-\zeta t+x & \zeta\tau t'+x'  & -x-x' \end{array}\right)\,.
\label{Q3_3x3}
\end{eqnarray}

The eigenvalues are given by:
\begin{eqnarray}
E^{(p)}(\tau,\zeta) & = & -(x+x')\pm\sqrt{A^{(\tau)}+\zeta B^{(\tau)}},\,\,
\nonumber\\
 & & \zeta(\tau t'-t)\,
\label{e03_tau_zeta}
\end{eqnarray}
with
\begin{eqnarray}
A^{(\tau)} & = & (x^2-xx'+x'^2)+(t^2+\tau tt'+t'^2)\,,\nonumber\\
B^{(\tau)} & = & -(2t+\tau t')x+(2\tau t'+t)x'\,.\nonumber\\
\label{e03_sb}
\end{eqnarray}
The 4 eigenstates corresponding to the eigenvalues $\zeta(\tau t'-t)$
which are independent of the spin couplings $J,J'$ have maximal
total spin $S=3/2$. The remaining ones have total spin $S=1/2$.

For the choices (\ref{tuning})(a,b) of plaquette parameters $(t,J,t',J')$
the ground state is given by one of the $J,J'$-dependent ($\tau=1$)-states
with energy:
\begin{eqnarray}
\hspace{0.6cm}E^{(p)}(\tau=1,\zeta) & =  & 
-\left(x+x'+\sqrt{A^{(1)}+\zeta B^{(1)}}\right)\,.\nonumber\\
\label{erg_Q3}
\end{eqnarray}
The ground state in the symmetric case $t'=t$, $J'=J$
\begin{eqnarray}
E^{(p)}(\tau,\zeta=\pm 1) & = & -J-\sqrt{\left(\frac{J}{2}\right)^2+3}
\end{eqnarray}
is twofold degenerate with respect to the quantum number $\zeta=\pm 1$.
This degeneracy is lifted in the asymmetric case (\ref{erg_Q3})
if the term $B^{(1)}$ (\ref{e03_sb}) is nonvanishing.

%

\item[e)]
$Q=4$, \,\,$N_c(4)=6$

We introduce $|m,n\rangle$ as $Q=4$-basis state with the two spin-up
particles ($+$) at plaquette positions $m$ and $n$.
Again, we use the definition that all creation operators of the
four electrons $2(+),2(-)$ 
act in order of increasing site number (cf. Fig. \ref{figa1})
on the vacuum $|0\rangle$ (e.g. $|1,3 \rangle\equiv c_{1\uparrow}^+
c_{2\downarrow}^+c_{3\uparrow}^+c_{4\downarrow}^+|0\rangle$).

In the basis:
\begin{eqnarray}
|1,\tau\rangle & = & |1,3\rangle + \tau|2,4\rangle\nonumber\\
|2,\tau\rangle & = & |1,2\rangle + \tau|3,4\rangle\\
|3,\tau\rangle & = & |1,4\rangle + \tau|2,3\rangle\nonumber
\end{eqnarray}
the $t-J$ Hamiltonian reduces to the matrix
\begin{eqnarray}
 & & \left(\begin{array}{ccc} -J & 0 & \Delta\tau J\\ 
0 & -J'  & \Delta\tau J'  \\
\Delta\tau J & \Delta\tau J'  & -(J+J')  \end{array}\right)\,.
\label{Q4_3x3}
\end{eqnarray}
with $\Delta\tau=(1+\tau)/2=1,0$.

The eigenvalues for $\tau=-1$ are simply given by:
\begin{eqnarray}
E^{(p)}(Q=4,-) & = & -J,\,-J',\,-(J+J'),
\end{eqnarray}
whereas the case $\tau=1$ yields:
\begin{eqnarray}
 & & E^{(p)}(Q=4,+)=\nonumber\\
 &  & -(J+J')\pm\sqrt{J^2+J'^2-JJ'},\,0\,.
\label{e04_s}
\end{eqnarray}
The ground state for all parameter values is uniquely given in
the $\tau=1$ sector.

\end{itemize}


\section{First order perturbation theory}
\label{appendix_b}          

\begin{figure}[ht!]
\centerline{\hspace{0.0cm}\includegraphics[width=5.0cm,angle=0]{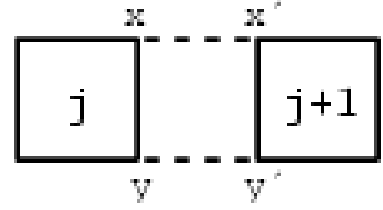}
\hspace{-0.0cm}}
\caption{Leg couplings ($x,x'$), ($y,y'$) linking the
neighbouring plaquettes $j$ and $j+1$ on the ladder.}
\label{figA1}
\end{figure}

To see the effects of first order perturbation theory in
the hopping parameter $t''$, we start from the transition
matrix elements:
\begin{eqnarray}
 & & \hspace{-0.0cm}t''\left\langle\prod_{j=1}^{N/4}\psi_0^{(p)}(Q'_j)\left|
\sum_{j=1}^{N/4}h_{j,j+1}(\alpha')\right|\prod_{j=1}^{N/4}\psi_0^{(p)}(Q_j)
\right\rangle
\nonumber\\
 & & \hspace{-0.0cm}=t''\sum_{j=1}^{N/4}A_{j,j+1}\left\langle\left.\psi_0^{(p)}(Q'_j)
\psi_0^{(p)}(Q'_{j+1})\right|h_{j,j+1}(\alpha')\Big|\right.\times\nonumber\\
 & & \hspace{3.5cm}\left.\left|\psi_0^{(p)}(Q_j)
\psi_0^{(p)}(Q_{j+1})\right.\right\rangle\nonumber\\
\end{eqnarray}
with
\begin{eqnarray}
A_{j,j+1} & = & \prod_{l\neq j,j+1}\delta_{Q'_l Q_l}
\end{eqnarray}
The interaction Hamiltonian $h_{j,j+1}(\alpha')$ between neighbouring
plaquettes is illustrated in Fig. \ref{figA1}. The hopping part
$h^{(t)}_{j,j+1}$ is active if the two links $\langle xx'\rangle$
$\langle yy'\rangle$ are occupied by one electron and one hole,
respectively. This means in terms of occupation numbers $n(x),n(x')$
on the sites $x,x'$ $n(x)=1,n(x')=0$ or $n(x)=0,n(x')=1$.
Therefore, the hopping term induces a charge exchange by one unit
\begin{eqnarray}
h^{(t)}_{j,j+1}:\quad (Q_j,Q_{j+1}) & \rightarrow & (Q_j-1,Q_{j+1}+1),
\nonumber\\
 & & \hspace{0.0cm}(Q_j+1,Q_{j+1}-1)
\label{QpQpp}
\end{eqnarray}
between neighbouring plaquettes. Note in particular, that charge exchange
by two and more units
\begin{eqnarray}
(0,2)\leftrightarrow (2,0) & \quad & (2,4)\leftrightarrow (4,2)
(Q_j+1,Q_{j+1}-1)\nonumber\\
\end{eqnarray}
is forbidden in first order perturbation theory.

Therefore the nonvanishing matrix elements:
\begin{eqnarray}
 & & \hspace{-0.0cm}\left\langle\psi_0^{(p)}(Q'_j)
\psi_0^{(p)}(Q'_{j+1})\Big|h_{j,j+1}(\alpha')
\Big|\psi_0^{(p)}(Q_j)
\psi_0^{(p)}(Q_{j+1})\right\rangle\nonumber\\
 & & =W(Q_j,Q_{j+1})\delta_{Q'_j,Q_j}\delta_{Q'_{j+1},Q_{j+1}}
\end{eqnarray}
are necessarily diagonal for the pairs of interest
\begin{eqnarray}
(Q_j,Q_{j+1}) & = & (0,2),(0,4),(2,4),(0,0),(2,2),(4,4)\,.\nonumber\\
\label{pairs}
\end{eqnarray}
They arise from the spin exchange part $h^{(J)}_{j,j+1}$ which
is active if both sites $x,x'$ (cf. Fig. \ref{figA1}) are occupied
with one electron:
\begin{eqnarray}
h^{(J)}_{j,j+1} & = & \left(2{\bf S}(x){\bf S}(x')+\frac{1}{2}\right)
n(x)n(x')\,.
\end{eqnarray}
Here ${\bf S}(x)$ and ${\bf S}(x')$ are spin operators at
sites $x$ and $x'$ and we get for the interaction
energies $W(Q_j,Q_{j+1})$ for plaquette pairs (\ref{pairs}):
\begin{eqnarray}
W(Q_j,Q_{j+1}) & = & -t''\alpha'\frac{Q_jQ_{j+1}}{32}
\label{W_Qj_Qj+1}
\end{eqnarray}
if $Q_j$ and $Q_{j+1}$ come from Eq.(\ref{pairs}).

\end{appendix}





\end{document}